\documentclass[a4paper]{jpconf}
\usepackage{amssymb,amsmath,bm,graphicx,textcomp,iopams,cite}
\begin{document}

\title{Collective impurity effects in the Heisenberg triangular antiferromagnet}

\author{V S Maryasin$^1$ and M E Zhitomirsky$^2$}
\address{$^1$
University Grenoble-Alps and INAC-SPSMS, F-38000 Grenoble, France}
\address{$^2$
Service de Physique Statistique, Magn\'etisme et Supraconductivit\'e,
INAC, CEA,\\ 17 rue des Martyrs, 38054 Grenoble Cedex 9, France}

\ead{vladimir.maryasin@cea.fr, mike.zhitomirsky@cea.fr}


\begin{abstract}
We theoretically investigate the Heisenberg antiferromagnet on a triangular lattice
doped with nonmagnetic impurities. Two nontrivial effects resulting
from collective impurity behavior are predicted. The first one is related to presence
of uncompensated magnetic moments localized near vacancies as revealed by
the low-temperature Curie tail in the magnetic susceptibility.
These moments exhibit an anomalous growth with the impurity concentration,
which we attribute to the clustering mechanism.
In an external magnetic field, impurities lead to an even more peculiar phenomenon lifting
the classical ground-state degeneracy in favor of the conical state. We analytically demonstrate that
vacancies spontaneously generate a positive biquadratic exchange, which is responsible
for the above degeneracy lifting.
\end{abstract}

\date{\today}

\section{Introduction}
Due to inherent competition of the principal interactions, frustrated magnets are susceptible
to effects at much lower energy scales that are determined by additional interactions,
fluctuations, or structural disorder. Though, the first two sources of degeneracy lifting have
been extensively studied in the literature, resulting in a wide range of theoretical models,
a detailed phenomenology of the impurity effects on frustrated magnets is lacking so far.
Most of the existing studies focus on a single impurity problem. These include magnetic
susceptibility of an impurity \cite{Sachdev99, Sachdev03, Sushkov03}, spin renormalization and
screening patterns around an impurity \cite{Hoglund07, Eggert07, Wollny11, Wollny12} and
partial relief of frustration \cite{Chen11}.

In this work we study the effect of dilution with nonmagnetic impurities
on the paradigmatic frustrated model of
the Heisenberg triangular-lattice antiferromagnet (TAFM) \cite{Lee84, Kawamura85}. We predict
two essentially collective impurity phenomena, which develop only at finite concentration of
defects $n_{\rm imp}$. The considered spin Hamiltonian is an isotropic classical antiferromagnet
on a triangular lattice:
\begin{equation}
\hat{\cal H} = J \sum_{\langle ij\rangle} \,\mathbf{S}_{i}\cdot\mathbf{S}_{j}(1-p_i)(1-p_j)
- {\bf H}\cdot \sum_{i}{\bf S}_i(1-p_i).
\label{H}
\end{equation}
Impurities are introduced via the parameter $p_i$, which is set to $p_i=0$ on regular sites
and to $p_i=1$ on impurity sites, hence, $p_i^2=p_i$. Impurities are assumed to be randomly distributed over
the lattice and $\sum_i p_i = N_{\rm imp}$.

Our first result is a nontrivial dependence of an effective impurity moment on the fraction of vacant
sites $n_{\rm imp}$. In noncollinear magnets even in the absence of external magnetic field
an impurity induces transverse local field on its neighbors and leads to the screening of magnetic moment of a missing spin.
Wollny \textit{et al.} \cite{Wollny11} found that in the Heisenberg TAFM the magnetic moment around a vacancy is equal
to $m_{\rm imp}^\circ = 0.039S$.

Here we report a substantial growth of the effective moment of a single impurity at finite concentrations
of vacancies, observed in our numerical simulations.
This growth is significant even at low concentrations $n \leqslant 1\%$, causing deviations from
independent impurity behavior with $m_{\rm imp} = m_{\rm imp}^\circ$.
We explain it by the effects of clustering of impurities.
The results of our numerical simulations, as well as the details of numerical methods, used in this
study are presented in the section \ref{sect:moment}.

The second effect of impurities considered in the present work is lifting of the continuous ground state
degeneracy of the Heisenberg TAFM in an external magnetic field.
This degeneracy gives rise to a rich phase diagram with various coplanar states \cite{Lee84, Kawamura85, Gvozdikova11}
that can be understood theoretically using the concept of order by disorder.
In our recent article \cite{Maryasin13} we have obtained numerically that vacancies doped into
TAFM  stabilize the conical ground state. Such a selection takes place because of a positive biquadratic exchange
produced by structural disorder. The concept of an effective biquadratic exchange generated
by quenched disorder in magnetic solids was first suggested by Slonczewski
in the context of magnetic multilayers \cite{Slonczewski91}.
In our previous work on an impure TAFM \cite{Maryasin13}
we have derived analytically a positive biquadratic exchange in a model with
weak bond randomness, see also  \cite{Maryasin14}.
In Sec.~\ref{sect:biq} we generalize the previous analytic results by showing that the effective biquadratic
exchange follows also from a weak site disorder.

\begin{figure}[t]
\center{\includegraphics[width=0.75\columnwidth, keepaspectratio]{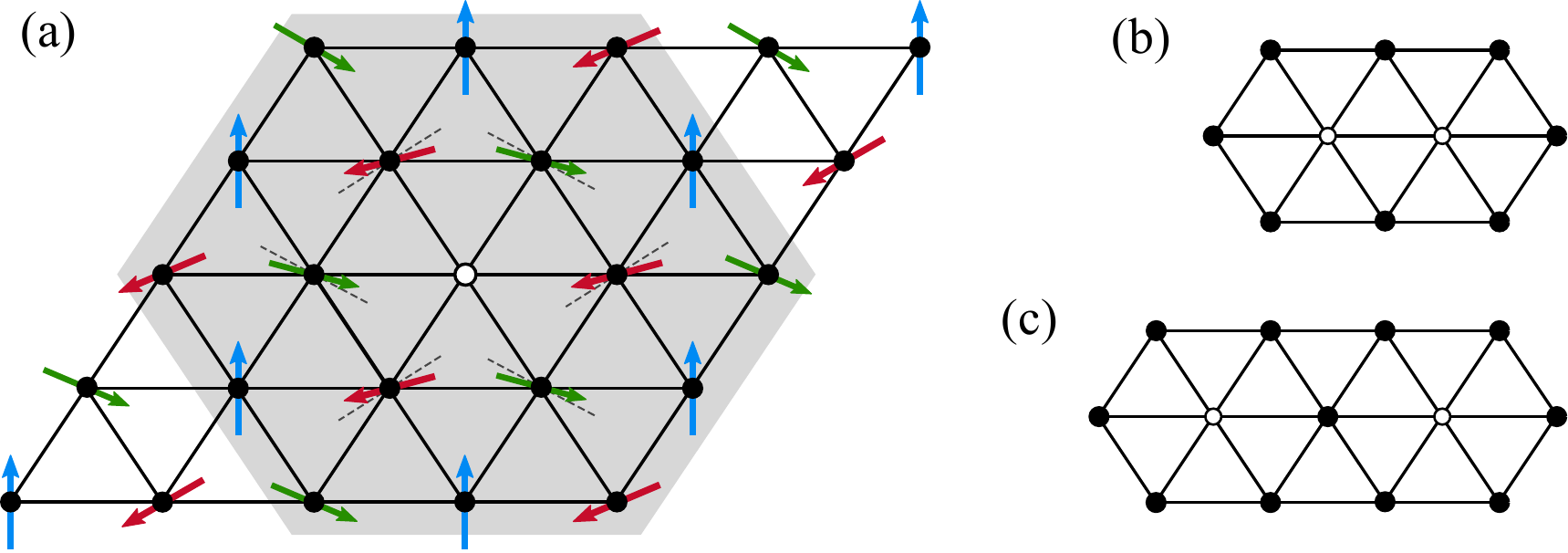}}
\caption{(a) An example of spin distortions around the impurity in the classical $120^\circ$ ordered state. Shaded area shows all nearest neighbours up to the order 3. (b), (c) Simple clusters of impurities with the strongest distortion of screening. Effective magnetic moment of these configurations equals to $m_{\rm imp}^{\circ\circ} = 0.11$ and $m_{\rm imp}^{\circ \cdot \circ} = 0.08$ respectively.}
\label{fig:conf}
\end{figure}

\section{Effective impurity moment}
\label{sect:moment}

A removed magnetic moment in an antiferromagnetic insulator induces a net magnetic moment
in the system. For noncollinear magnetic structures the moment is, however, screened by a spin
texture resulting from the canting of the surrounding spins \cite{Eggert07, Wollny11, Wollny12}.
Such a spin canting  of only nearest neighbor spins is illustrated in Fig.~\ref{fig:conf}(a).
As a result,  an impurity moment acquires a non-universal fractional value, which depends on
the system details. Wollny and collaborators \cite{Wollny11} have shown that a vacancy
in the classical Heisenberg TAFM becomes slightly overcompensated, i.e.\ at $T=0$ the net
magnetic moment is equal to $m_{\rm imp}^\circ = 0.039S$ and has the {\it same} direction as a
missing spin. At finite temperatures in the absence of long range order this purely classical
moment is  free to rotate in spin space leading to a Curie-like paramagnetic divergence of
the magnetic susceptibility at $T \rightarrow 0$. In a system with a small but finite concentration of
impurities, the impurity contributions sum up to give
\begin{equation}
\chi(T) = \frac{N_{\rm imp}m_{\rm imp}^2}{3T} + O(1).
\label{chi}
\end{equation}

In this section we consider dependence of the effective impurity moment $m_{\rm imp}$ on the
impurity concentration $n_{\rm imp}$.
We have numerically investigated a finite system with a fixed finite concentration of vacancies,
measured $m_{\rm imp}$ and studied how the impurity screening is modified at finite $n_{\rm imp}$.

First we report results of Monte Carlo simulations of the classical nearest-neighbor
Heisenberg antiferromagnet (\ref{H}) on rhombohedral $L \times L$ clusters with finite concentration of static vacancies.
Basically, we use the same algorithm, as in the previous works \cite{Gvozdikova11, Maryasin13}.
We found that cluster size has little effect on bulk thermodynamic quantities at $T \rightarrow 0$ and, therefore,
used moderate cluster size $L=90$ in all runs for this work, except for the cases of small amount of impurities
$n_{\rm imp} < 0.01$, where larger clusters are needed for better statistics.

\begin{figure}[t]
\center{\includegraphics[width=0.56\columnwidth, keepaspectratio]{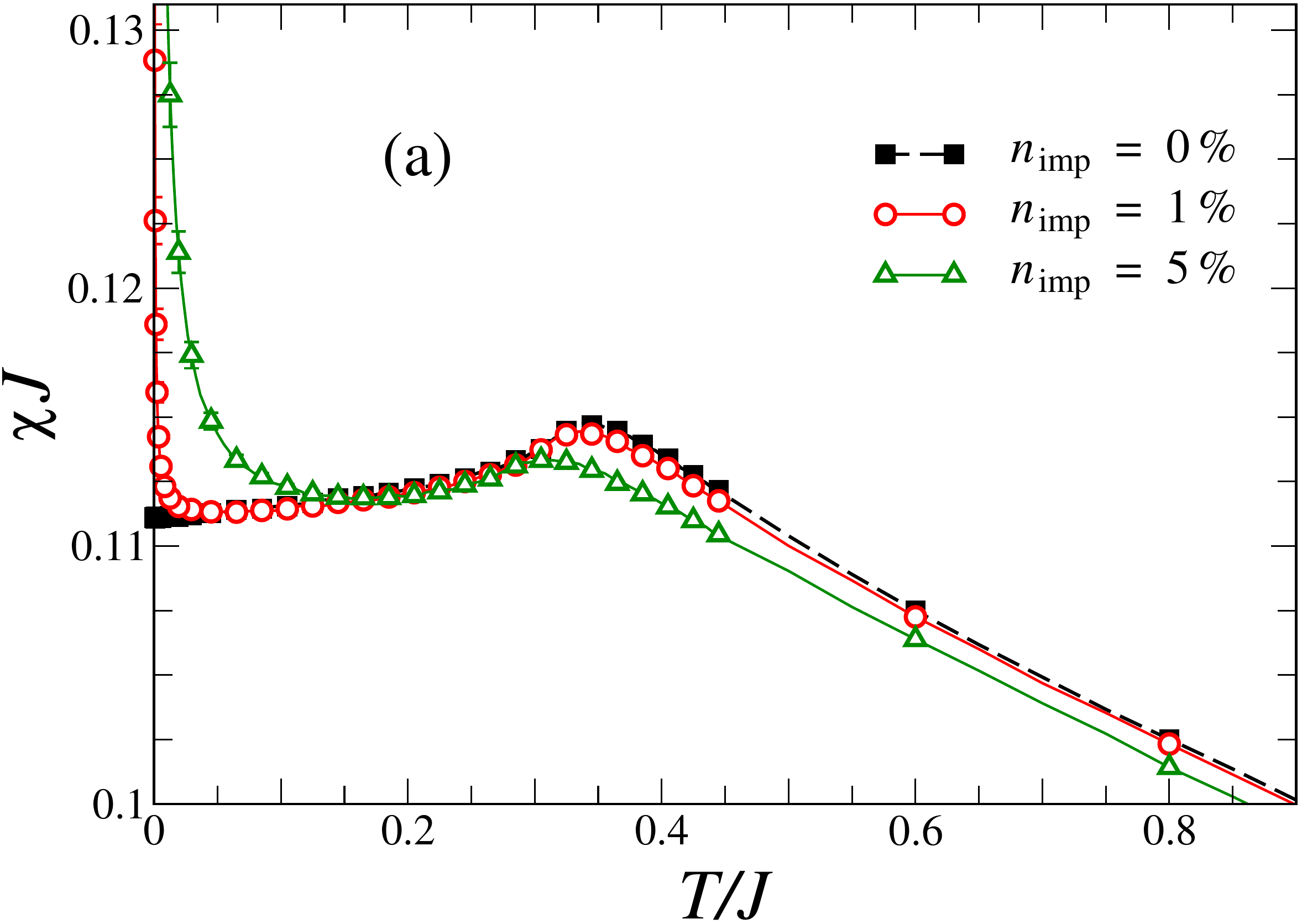}}
\vspace{10pt}
\center{\includegraphics[width=0.55\columnwidth, keepaspectratio]{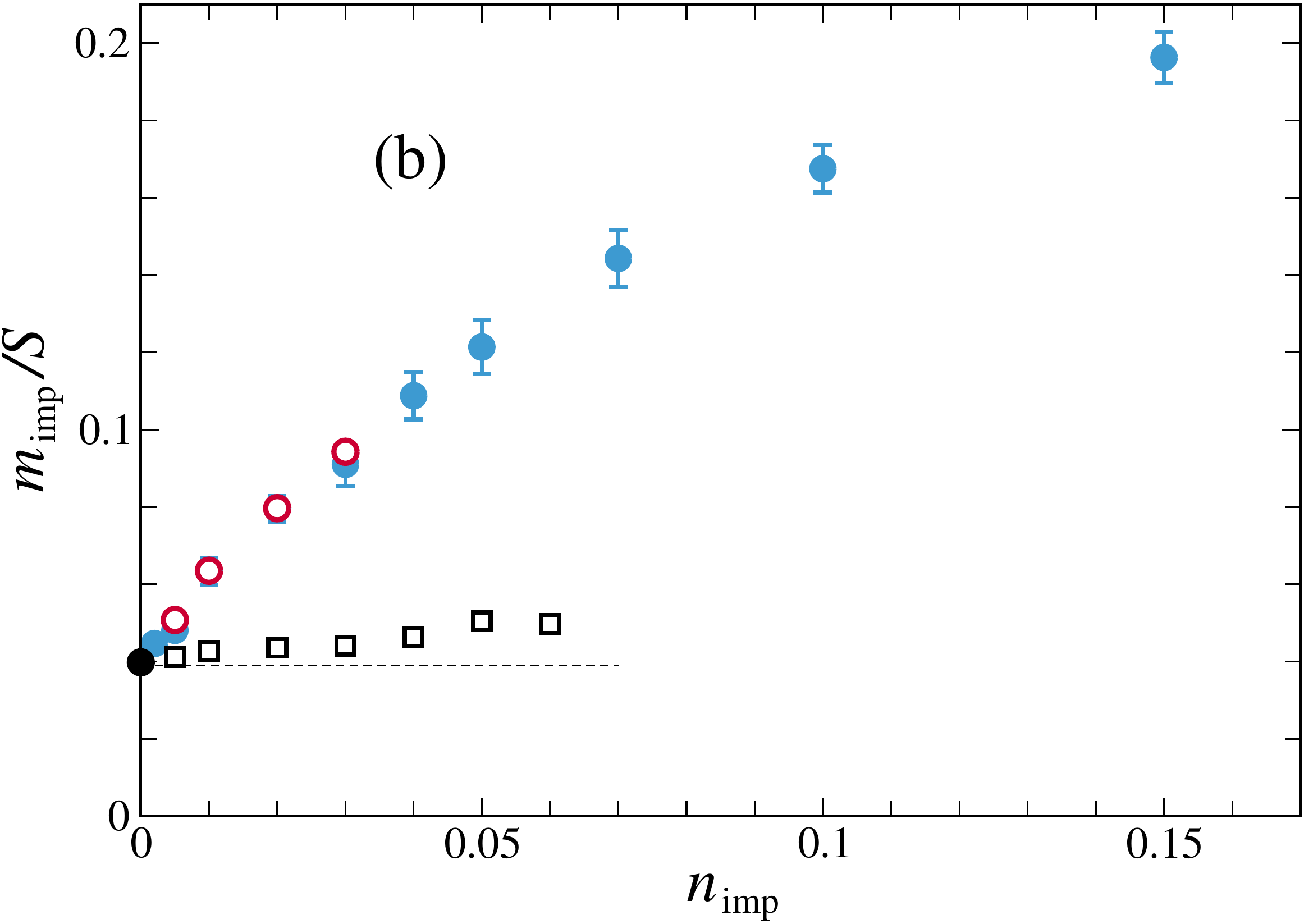}}
\caption{(a) Magnetic susceptibility for different concentrations of vacancies, from our Monte Carlo simulations. The Curie-like singularity at $T \rightarrow 0$ gives the value of effective impurity moment $m_{\rm imp}$. (b) Growth of effective vacancy moment with vacancy concentration obtained using susceptibility interpolation (full circles) and direct measurement at the ground state (open circles). Open squares correspond to simulation of the system with impurities, distributed at a distance from each other. Dashed line and a big marker at $n_{\rm imp} = 0$ show $m_{\rm imp}^\circ$ obtained in the work \cite{Wollny11}.}
\label{fig:m_imp}
\end{figure}

Figure \ref{fig:m_imp}(a) shows the uniform magnetic susceptibility $\chi(T)$ normalized per spin
obtained from the Monte Carlo
simulations of the TAFM  with and without impurities as
\begin{equation}
\chi = \frac{1}{3TL^2}\Bigl\langle \Bigl(\sum_i \mathbf{S}_i \Bigr)^2\Bigr\rangle.
\end{equation}
The main difference between the curves is the emergent $1/T$ divergence of $\chi$ at low temperatures, which becomes
stronger with increasing $n_{\rm imp}$. This upturn may look counterintuitive as no extra magnetic moments are brought
into the system. Note that our Monte Carlo results for $\chi(T)$ closely resemble the susceptibility
data measured for nominally pure TAFM materials,  for example,  for $\rm LuMnO_3$ \cite{Lewtas10}.

We associate an average magnetic moment $m_{\rm imp}$ with every impurity and interpolate
the susceptibility curves at $T \rightarrow 0$ with Eq.~(\ref{chi}) to determine its value.
In addition, we obtain independent results for the impurity moments $m _{\rm{imp}}$ by another method:
direct calculation of
\begin{equation}
m_{\rm{imp}}^2=\frac{1}{N_{\rm imp}} \Bigl( \sum_i \mathbf{S}_i \Bigr)^2
\end{equation}
in the classical ground state at zero temperature.
The algorithm of the search of the ground state has been described in details
in the works \cite{Maryasin13, Maryasin14}. Note that $m _{\rm{imp}}$, obtained by both methods
was averaged over at least 200 random impurity configurations.

Figure \ref{fig:m_imp}(b) presents the main result of this section: a nontrivial growth of
single impurity moment with concentration.
Values, obtained by the two methods (displayed by full and open circles respectively) match perfectly
and in the following we do not make the difference between the two methods.
The growth is observed even at small $n _{\rm{imp}} \sim 1\%$, which is somewhat surprising
as at such weak dilution one may expect a nearly independent impurity behavior with $m _{\rm{imp}}=m_{\rm imp}^\circ$.
Indeed, Fig.~\ref{fig:m_imp} (b) shows that $m _{\rm{imp}}$ is significantly renormalized from
the single impurity value, even at the lowest considered $n _{\rm{imp}} = 0.2\%$, which is
comparable to the case of a single impurity on a cluster with $L=12$, studied in the work of
Wollny \textit{et al.} \cite{Wollny11}.

One possible explanation of growing $m _{\rm{imp}}$ is an assumption that screening clouds from different
impurity sites interact, and self-average. According to Ref.~\cite{Wollny11, Luscher05} readjustment angles of spins,
surrounding an impurity decay as $\delta \Theta(r) \sim 1/r^3$ at long distances. For a finite concentration
of impurities one may expect cut off of individual screening clouds at average distances
$R \sim n _{\rm{imp}}^{-1/2}$. It leads to the modification of screening, consistent with the square root dependence on vacancy fraction:
\begin{equation}
\delta m \sim \int \Theta(r)\vert\mathbf{l}(r)\vert d^2 r \sim n _{\rm{imp}}^{1/2}.
\end{equation}
Here $\mathbf{l}(r)$ - is a vector, pointing in the direction spin distortion, and results only in a prefactor, which is omitted. However, the measured $m_{\rm imp}$ (fig. \ref{fig:m_imp} (b)) does not fit to this procedure, and therefore demands for different explanation of the observed dependence.

We ascribe growth of impurity moment to the effects that are quadratic in $n _{\rm{imp}}$.
Indeed, an individual impurity moment is strongly screened to a very small value $m_{\rm imp}^\circ$.
Hence, one is forced to consider statistically rare cases of two impurities occupying nearby sites, see Figs.~\ref{fig:conf}(b) and (c).
If such impurity configurations have moments, which are not very well screened and which are significantly larger
than $m_{\rm imp}^\circ$, their impact on the net magnetic moment and the low-temperature susceptibility may
be quite significant.

We measured $m_{\rm imp}$ for a few simple vacancy configurations and show in Figs.~\ref{fig:conf}(b) and (c)
two impurity clusters with the largest values of the effective magnetic moment. The calculations yield
$m_{\rm imp}^{\circ\circ} = 0.11$ and $m_{\rm imp}^{\circ \cdot \circ} = 0.08$ respectively, the result,
that is several times larger, than $m_{\rm imp}^\circ$. Along with the large coordination number of
the triangular lattice, it overcomes the small statistical weight $P \sim n_{\rm imp}^2$ of these configurations.

In addition, we have performed similar numerical simulations restricting vacancies from being placed
within the first three neighbors from each other. The respective exclusion region is
is shown by the shaded area in Fig.~\ref{fig:conf}(a). In the first place, such a computation serves
to verify the above hypothesis; second it may also model weak correlations in the structural disorder, which
develop in real solids due to elastic tension etc.
The results are plotted in Fig.~\ref{fig:m_imp}(b) with open square markers.
They demonstrate only a slight growth of $m_{\rm imp}$ from $m_{\rm imp}^\circ$. Therefore, at distances
exceeding 2--3 lattice spacings, impurities only barely interact and behave completely individually.
These results strongly support the above explanation of the growth of impurity moment due to clustering of vacancies.

\section{Disorder induced biquadratic exchange}
\label{sect:biq}

In this section we elaborate on the mechanism of the degeneracy lifting
in the TAFM as well as in a wide class of frustrated models produced by nonmagnetic impurities.
In finite magnetic fields $0 < H < H_{\rm sat}$ the TAFM exhibits continuous degeneracy
of ground states obeying the classical constraint ${\bf S}_\triangle = {\bf H}/3J$.
Thermal  \cite{Lee84, Kawamura85} and quantum \cite{Chubukov91}
order by disorder are responsible for selecting the most collinear spin configurations.
These include the coplanar `Y' and `2:1' states below and above $\frac{1}{3}H_{\rm sat}$, respectively,
and the collinear up-up-down state at the $1/3$ magnetization plateau.
The resulting complex phase diagram of the classical Heisenberg TAFM was obtained
from the Monte Carlo simulations in \cite{Gvozdikova11}.
At the phenomenological level, such a selection can be straightforwardly explained
by appearance of an effective biquadratic exchange with negative sign: $-({\bf S}_i\cdot{\bf S}_j)^2$.
Such term was indeed obtained in the lowest order of the real-space perturbation
theory for quantum \cite{Long89,Heinila93} and thermal \cite{Canals04} fluctuations.
Below we present the opposite effect of \textit{order by structural disorder}. We show that a small
finite concentration of weak vacancies may be described by a positive biquadratic exchange
$+({\bf S}_i\cdot{\bf S}_j)^2$ in an effective spin Hamiltonian. Therefore, structural disorder stabilizes the least collinear
spin configurations, which in the case of the TAFM correspond to conical states, and competes with conventional
order by disorder mechanism.

For analytic derivation we need to modify the model for impurities, introduced in Eq.~(\ref{H}).
As was discussed in the previous section, a true vacancy produces a strong long-range distortion
of the surrounding spin structure. Therefore, below we restrict ourselves to a model of
weak classical impurities with reduced spin length $\mathbf{S}_{i}(1-\varepsilon p_i)$, and $\varepsilon \ll 1$.
This approximation is analogous to restricting the impact of impurities to its nearest neighbors only,
as illustrated in Fig.~\ref{fig:conf}(a). Numerical results of our previous work \cite{Maryasin13} demonstrate
that this approximation does not affect the conclusion. Therefore,
we expect the state selection produced by real vacancies to be even more robust.

In the following we follow the ideas of the real-space perturbation theory \cite{Long89, Heinila93, Canals04}.
We start at $T=0$ with an arbitrary classical ground state of the pure system and express the Hamiltonian in terms of
small spin deviations, caused by impurities. Then, we collect all single-site terms and find a
static distortion of the equilibrium magnetic structure from a simple minimization procedure.
The remaining terms, that represent interaction of perturbations, will be neglected, as they produce higher-order
corrections \cite{Maryasin14}.

An arbitrary classical ground state of the pure system is characterized by a set of angles $\theta_{ij}$
between neighboring spins. Performing transformation to a  local rotating frame we obtain
\begin{eqnarray}
\hat{\cal H} = J\sum_{\langle ij\rangle} \bigl[  S_i^y S_j^y + \cos\theta_{ij}
\bigl(S_i^z S_j^z + S_i^x S_j^x \bigr)
+ \sin\theta_{ij} \bigl(S_i^z S_j^x - S_i^x S_j^z \bigr) \bigr]\bigl[1-\varepsilon (p_i + p_j)\bigr]
+ \hat{\cal H}_{\rm Z}.
\label{H1}
\end{eqnarray}
The local $\hat{\mathbf{z}}_{i}$ axes are chosen along the spin direction on each site, whereas the $\hat{\mathbf{x}}_{i}$ axes
lie within the plane, formed by a pair of spins $\mathbf{S}_i$ and $\mathbf{S}_j$.
Impurities produce distortions of the spin structure, generating small transverse components components of spins $S^x$ and $S^y$
that correspond to in plane and out of plane spin deviations. Then one can write
\begin{equation}
S^z = \sqrt{1-(S^x)^2-(S^y)^2} \approx 1-\frac{1}{2}[(S^x)^2)+(S^y)^2]
\label{Sz}
\end{equation}
All terms arising from interaction of spin deviations on adjacent sites are neglected.
Therefore the leading ground-state energy correction  comes from the contribution that is linear in in-plane
spin deviations $S^x$ and disorder $\varepsilon$
\begin{equation}
V_1 = J\varepsilon \sum_{\langle ij \rangle}\sin \theta_{ij}\left(S_i^xp_j - S_j^xp_i \right).
\label{V}
\end{equation}
Physically, terms linear in spin deviations appear due to a local relief of magnetic frustration
caused by impurities. Excluding the out of plane spin components $S^y$ that do not enter in $V_1$,
and we are left with
\begin{eqnarray}
\hat{\cal H} = \sum_i \Biggl[ \frac{1}{2}H_{\rm loc} (S_i^x)^2 + J\varepsilon S_i^x \sum_{j=1}^{6} \sin \theta_{ij} p_j \Biggr].
\label{Hmin}
\end{eqnarray}
Here $H_{\rm loc}=|\partial E_{\rm g.s.}/\partial {\bf S}_i|= 3J$ is a local field, which is the same on all 
three sublattices of the pure TAFM irrespective of the applied field $H<H_{\rm sat}$.
Minimizing this quadratic form, we obtain the correction to the ground state energy, generated by disorder.
Note that throughout the derivation, we omitted all constant terms, concentrating only on the configuration-dependent correction.
\begin{equation}
\Delta E = - \frac{\varepsilon^2 n_{\rm{imp}}}{H_{\rm loc}} \sum_{\langle ij \rangle} \sin^2 \theta_{ij} =
 \frac{\varepsilon^2 n_{\rm{imp}}}{3J} \sum_{\langle ij \rangle} (\mathbf{S}_i \cdot \mathbf{S}_j)^2 .
\label{DE}
\end{equation}
Similarly to the case of bond disorder \cite{Maryasin13} we obtained, that realistic impurities in the TAFM
favor the conical spin state with the smallest values of $\cos \theta_{ij}$, as spin relaxation from
the least collinear configurations produces the largest energy gain.
The realistic model, given by equation (\ref{H}) produces the similar positive biquadratic exchange term
as random bond disorder but in principle suits more to describe numerical simulations of Ref.~\cite{Maryasin13}
as well as experiments on doped frustrated systems.
Finally, we note that the above derivation stays completely intact for the TAFM with a planar anisotropy
considered in the very first work \cite{Lee84}.
In this case the conical states are suppressed by the $XXZ$ anisotropy in the spin Hamiltonian.
Still, the classical degeneracy is present within the manifold of the coplanar states \cite{Lee84}.
In this case a positive biquadratic exchange will stabilize an anti-Y sate in the whole
range of magnetic fields \cite{Maryasin13}.

\section{Conclusions}
In this work we have described two effects, which appear in the frustrated Heisenberg triangular
antiferromagnet with a small but finite density of nonmagnetic impurities. First, we demonstrate that
an effective impurity moment revealed in the paramagnetic Curie tail of the magnetic susceptibility
exhibits a substantial growth with the impurity concentration. We attribute this growth to
an anomalously small value of the magnetic moment of an isolated vacancy in the TAFM, and, as a consequence,
to significance of correlated impurity effects $\sim n_{\rm imp}^2$.
This effect may be quantitatively different in other noncollinear helical antiferromagnets
with not so small values of vacancy moments.
Second, in the model of `weak' impurities we have analytically derived
an effective positive biquadratic exchange generated by the structural disorder.
Such an interaction term competes with the effect of thermal and quantum fluctuations
and plays a very important role in the phenomenology of the complex phase diagram of the
doped TAFM \cite{Maryasin13}. With minor modifications, mainly in the value of $H_{\rm loc}$,
the presented derivation holds for impurities in other geometrically frustrated 
magnets.

\end{document}